\documentclass[12pt]{article}

\usepackage{graphicx}
\usepackage{a4}

\usepackage{fancybox}
\usepackage{epsfig}
\usepackage{color}

\usepackage{amsfonts}
\usepackage{mathrsfs}
\usepackage{amssymb}
\usepackage{amsmath}

\usepackage{dcolumn}
\usepackage{bm}

\def\pa{\partial}

\def\al{\alpha}

\def\de{\delta}
\def\ep{\epsilon}

\def\th{\theta}

\def\la{\lambda}

\def\om{\omega}

\def\Ga{\Gamma}

\def\La{\Lambda}
\def\Si{\Sigma}

\def\Om{\Omega}

\newcommand{\ben}{\begin{equation}}
\newcommand{\een}{\end{equation}}
\newcommand{\bea}{\begin{eqnarray}}
\newcommand{\eea}{\end{eqnarray}}
\newcommand{\ba}{\begin{array}}
\newcommand{\ea}{\end{array}}
\newcommand{\bit}{\begin{itemize}}
\newcommand{\eit}{\end{itemize}}

\newcommand{\vs}[1]{\vspace{#1 mm}}

\newcommand{\dsl}{\pa \kern-0.5em /}

\begin{document}

\topmargin 0pt \oddsidemargin 0mm

%\begin{flushright}

%USTC-ICTS-11-10 \\

%\end{flushright}

\vspace{2mm}

\begin{center}

{\Large \bf Force-free magnetosphere on near-horizon geometry of
near-extreme Kerr black holes}
%{\Large \bf Force-free magnetosphere on near-NHEK}

\vs{10}

 {\large Huiquan Li \footnote{E-mail: lhq@ynao.ac.cn}, Cong Yu, Jiancheng Wang and Zhaoyi Xu}

\vspace{6mm}

{\em

Yunnan Observatories, Chinese Academy of Sciences, \\
650011 Kunming, China

Key Laboratory for the Structure and Evolution of Celestial Objects,
\\ Chinese Academy of Sciences, 650011 Kunming, China}

\end{center}

\vs{9}

\begin{abstract}
We study force-free magnetospheres in the Blandford-Znajek process
from rapidly rotating black holes by adopting the near-horizon
geometry of near-extreme Kerr black holes (near-NHEK). It is shown
that the Znajek regularity condition on the horizon can be directly
derived from the resulting stream equation. In terms of the
condition, we split the full stream equation into two separate
equations. Approximate solutions around the rotation axis are
derived. They are found to be consistent with previous solutions
obtained in the asymptotic region. The solutions indicate energy and
angular-momentum extraction from the hole.
\end{abstract}

%\textit{Keywords:}

%\textit{PACS:}

\section{Introduction}
\label{sec:introduction}
%%%%%%%%%%%%%%%%%%%%%%%%%%%%%%%%%%%%%%%%%%%%%%%%%%%%%%%%%%%%%%%%%%%%%%%%%%%%

Black hole magnetospheres as well as many high-energy astrophysical
objects, for instance, relativistic jets in active galactic nuclei
and gamma-ray bursts and probably ultra-strongly magnetized neutron
stars (magnetars) involve relativistically magnetically dominated
plasma. Under such circumstances, magnetic fields play crucial roles
in the dynamics of these astrophysical scenarios, which can drive
powerful winds/jets from these astrophysical objects. In addition,
the magnetic dissipation can also give rise to remarkable
non-thermal emissions in the these high-energy astrophysical
phenomena. It is widely accepted that, in these objects, the
magnetic energy density conspicuously exceeds the thermal and rest
mass energy density of particles. The force-free electrodynamics
behave well in such extreme magnetically dominated scenarios as the
less important terms, such as the inertia and pressure, are entirely
ignored. In the framework of force-free electrodynamics, Blandford
and Znajeck (BZ) \cite{Blandford:1977ds} proposed that the rotation
energy of a kerr black hole could be extracted via the action of
force free electromagnetic fields, in the form of Poynting flux via
magnetic field lines penetrating the central black hole. The tapping
of the rotational energy form spinning black hole via magnetized
plasma appears to be the most astrophysically promising mechanism to
exploit the black hole energy. Relativistic jets in active galactic
nuclei, galactic microquasars, and gamma ray bursts may well be
driven by the BZ mechanism.

Exploring solutions in this configuration is crucial for examining
the energy extraction process and checking the efficiency of this
mechanism. However, very few analytical solutions have been obtained
so far (a collection of available solutions is summarised in
\cite{Gralla:2014yja} with modern treatments). In particular, it is
hard to derive the solutions in the case of rapidly rotating holes.
A key reason is that little information can be extracted from the
single stream equation, accompanied with a regularity condition on
the horizon \cite{Znajek:1977unknown}. In the original work
\cite{Blandford:1977ds}, split monopole and paraboloidal solutions
for slowly rotating black holes are obtained with expansion on the
angular momentum of the hole to the leading order. But, this
perturbative approach involving higher order corrections seems
unsuccessful in searching for solutions for rapidly rotating back
holes \cite{Tanabe:2008wm}. Nevertheless, a set of exact solutions
for arbitrary angular momentum are obtained in the far-field region
by Menon and Dermer (MD solutions) \cite{Menon:2005va,Menon:2005mg},
which indeed implies the formation of collimated outflow or jet
\cite{Menon:2011zu}. Based on previous solutions, generlised
solutions to time-dependent and non-axisymmetric case are obtained
by assuming the four-current to be along one of the null tetrads
\cite{Brennan:2013jla}.

In theoretical physics, extreme and near-extreme black holes attract
plenty of attention in past years, due to the proposal of the
AdS/CFT correspondence. For extremely rapidly rotating black holes,
this specific dual description is the Kerr/CFT correspondence
\cite{Guica:2008mu,Castro:2009jf,Bredberg:2011hp}, constructed on
the near-horizon geometry of (near-)extreme Kerr black holes, namely
(near-)NHEK \cite{Bardeen:1999px,Castro:2009jf}. In this paper, we
shall adopt the near-NHEK, replacing the full Kerr metric, to study
the force-free magnetosphere around near-extreme Kerr black holes.
The near-NHEK should be very suitable for seeking for solutions in
this system. First, it has a very simple form with precise
symmetries, so that we can simplify the equations and study them
analytically. Second, in the near-horizon region, we can take full
advantage of the constraint condition on the horizon. Hopefully, the
solution obtained on near-NHEK may provide us clues for searching
for solutions in the full Kerr background.

Meanwhile, the astrophysical processes, including the BZ process
near rapidly rotating holes, may also provide an arena for testing
the Kerr/CFT correspondence. Recently, there appear the works
\cite{Porfyriadis:2014fja,Hadar:2014dpa} along this line, where the
authors studied the gravity waves produced from massive objects
orbiting around an (near-)extreme Kerr on (near-)NHEK in the context
of Kerr/CFT. In \cite{Wang:2014vza}, the dual CFT aspects of the
magnetosphere around slow-rotating balck holes on AdS space has been
discussed.

Here, we try to adopt the near-NHEK to explore analytical solutions
of force-free magnetospheres near rapidly rotating holes. The paper
is organised as follows. In Section.\ \ref{sec:adsdescription}, we
present the near-NHEK and the expressions of EM fields on it. In
Section.\ \ref{sec:freemag}, we give the stream equation of
force-free magnetospheres on near-NHEK and then seek for solutions
to the equation. In Section.\ \ref{sec:extraction}, we show that it
is possible to form a outward flux from the derived solution. A
brief summary is made in the last section.

During the proceedings of this work, the works
\cite{Lupsasca:2014pfa,Lupsasca:2014hua} appear, which focus on
magnetospheres on exact NHEK by taking advantage of the conformal
symmetries of the geometry.

%%%%%%%%%%%%%%%%%%%%%%%%%%%%%%%%%%%%%%%%%%%%%%%%%%%%%%%%%%%%%%%%%%%%%%%%%%%%
\section{EM fields on the near-NHEK}
\label{sec:adsdescription}
%%%%%%%%%%%%%%%%%%%%%%%%%%%%%%%%%%%%%%%%%%%%%%%%%%%%%%%%%%%%%%%%%%%%%%%%%%%%

\subsection{The near-NHEK}

Consider Kerr black holes on the Boyer-Lindquist (BL) coordinates
\begin{equation}\label{e:Kerr}
 ds^2=\frac{\triangle}{\Si}(d\hat{t}-a\sin^2\th d\hat{\phi})^2
-\frac{\Si}{\triangle}dr^2-\Si d\th^2-\frac{\sin^2\th}
{\Si}((r^2+a^2)d\hat{\phi}-ad\hat{t})^2,
\end{equation}
where
\begin{equation}
 \triangle=(r-r_+)(r-r_-), \textrm{ }\textrm{ }\textrm{ }
r_\pm=M\pm\sqrt{M^2-a^2} \nonumber
\end{equation}
\begin{equation}
 \Si=r^2+a^2\cos^2\th, \textrm{ }\textrm{ }\textrm{ }
a=\frac{J}{M}. \nonumber
\end{equation}
$r_\pm$ are the radii of the inner and outer horizons, respectively,
and $J$ is the angular momentum of the black hole. The determinant
of the metric is $\sqrt{-\hat{g}}=\Si\sin\th$.

We adopt the following notations and coordinate transformations:
\begin{equation}\label{e:Kerrred11}
 r=r_+(1+\de), \textrm{ }\textrm{ }\textrm{ }
\de=\la u, \textrm{ }\textrm{ }\textrm{ }
\hat{t}=\frac{2r_+\bar{t}}{\la},
\end{equation}
\begin{equation}\label{e:Kerrred12}
 \hat{\phi}=\phi+\frac{\bar{t}}{\la\al},
\textrm{ }\textrm{ }\textrm{ } \al=\frac{a}{r_+}, \textrm{ }\textrm{
}\textrm{ } \la\ep=\frac{r_+-r_-}{r_+}=1-\al^2.
\end{equation}
To get the last equation, we have used the relation: $r_+r_-=a^2$.
Taking the value of $\la$ to be very small (this is also the
near-extreme case with $\al\rightarrow1$ for finite and fixed
$\ep$), we have the near-horizon geometry of (\ref{e:Kerr})
\begin{equation}\label{e:Kerr2}
 ds^2=2r_+^2\Ga(\th)^2\left[u(u+\ep)d\bar{t}^2-
\frac{du^2}{u(u+\ep)}-d\th^2-f(\th)^2\left(d\phi
+\frac{1}{2}(\ep+2u)d\bar{t}\right)^2\right],
\end{equation}
where
\begin{equation}\label{e:}
 \Ga(\th)^2=\frac{1}{2}(1+\cos^2\th), \textrm{ }\textrm{
}\textrm{ } f(\th)=\frac{\sin\th}{\Ga^2}.
\end{equation}
This near-horizon geometry for each fixed $\th$ can be viewed as a
warped $AdS_3$ space with the $SL(2,R)\times U(1)$ isometry group.
The geometry has been derived in \cite{Castro:2009jf} via different
coordinate redefinitions.

For later use, we further employ the following redefinitions
\begin{equation}\label{e:Kerrred3}
 U=\ep\sinh^2\frac{\rho}{2}, \textrm{ }\textrm{ }\textrm{ }
\bar{t}=\frac{2t}{\ep}.
\end{equation}
The metric simply becomes
\begin{equation}\label{e:Kerr3}
 ds^2=2r_+^2\Ga(\th)^2\left[\sinh^2\rho dt^2-d\rho^2-d\th^2-
f(\th)^2(d\phi+\cosh\rho dt)^2\right],
\end{equation}
Towards the extreme limit, $\rho$ ranges in $0\leq\rho<\infty$. The
event horizon is located at $\rho=0$. The determinant of the metric
becomes
\begin{equation}\label{e:}
 \sqrt{-g}=4r_+^4\Ga^2\sin\th\sinh\rho.
\end{equation}

Around the polar axis $\th\sim0$ and $\sim\pi$, we have
\begin{equation}\label{e:}
 \Ga(\th)^2\simeq1, \textrm{ }\textrm{
}\textrm{ } f(\th)\simeq\sin\th.
\end{equation}
In this case, the near-NHEK in $0\leq\rho\ll1$ is very similar to
the near-horizon geometry of Schwarzschild black holes (i.e., the
Rindler $\times$ sphere), except for a transformation
$d\phi\rightarrow d\phi+\cosh\rho dt$.

\subsection{EM fields}

The null tetrads in the Carter frame on the near-NHEK are derived
\begin{equation}\label{e:Cartet1}
 l^\mu=\frac{1}{\sqrt{2}r_+\Ga}\left(\frac{1}{\sinh\rho},
1,0,-\coth\rho\right),
\end{equation}
\begin{equation}\label{e:Cartet2}
 n^\mu=\frac{1}{\sqrt{2}r_+\Ga}\left(\frac{1}{\sinh\rho},
-1,0,-\coth\rho\right),
\end{equation}
\begin{equation}\label{e:Cartet3}
 m^\mu=\frac{1}{\sqrt{2}r_+\Ga}\left(0,0,
1,\frac{i}{f}\right).
\end{equation}
The tetrads are related to the electromagnetic (EM) field strength
via the scalars %(see e.g., [Z77])
\begin{equation}\label{e:}
 \phi_1=\frac{1}{2}F_{\mu\nu}(l^\mu n^\nu+\bar{m}^\mu m^\nu),
\nonumber
\end{equation}
\begin{equation}\label{e:}
 \phi_0=F_{\mu\nu}l^\mu m^\nu, \textrm{ }\textrm{ }\textrm{ }
\phi_2=F_{\mu\nu}\bar{m}^\mu n^\nu, \nonumber
\end{equation}
where $E_1+iB_1=2\phi_1$, $E_2+iB_2=\phi_2-\phi_0$ and
$E_3+iB_3=i(\phi_2+\phi_0)$. Hence, the EM fields near the
near-extreme Kerr are given by:
\begin{equation}\label{e:}
 E_\rho=\frac{-1}{r_+^2\Ga^2\sinh\rho}(F_{t\rho}+\cosh\rho
F_{\rho\phi}), \textrm{ }\textrm{ }\textrm{ }
B_\rho=\frac{1}{r_+^2\Ga^2f}F_{\th\phi},
\end{equation}
\begin{equation}\label{e:}
 E_\th=\frac{-1}{r_+^2\Ga^2\sinh\rho}(F_{t\th}+\cosh\rho
F_{\th\phi}), \textrm{ }\textrm{ }\textrm{ }
B_\th=\frac{-1}{r_+^2\Ga^2f}F_{\rho\phi},
\end{equation}
\begin{equation}\label{e:}
 E_\phi=\frac{-1}{r_+^2\Ga^2f\sinh\rho}F_{t\phi},
\textrm{ }\textrm{ }\textrm{ }
B_\phi=\frac{1}{r_+^2\Ga^2}F_{\rho\th}.
\end{equation}
Some components may be singular on the horizon since the tetrads
(\ref{e:Cartet1})-(\ref{e:Cartet3}) are singular on the horizon in
this frame. But if we transform the tetrads into the ones in the
ingoing frame by null rotation, the tetrads can be finite on the
horizon.

%%%%%%%%%%%%%%%%%%%%%%%%%%%%%%%%%%%%%%%%%%%%%%%%%%%%%%%%%%%%%%%%%%%%%%%%%%%%
\section{The force-free magnetosphere}
\label{sec:freemag}
%%%%%%%%%%%%%%%%%%%%%%%%%%%%%%%%%%%%%%%%%%%%%%%%%%%%%%%%%%%%%%%%%%%%%%%%%%%%

In addition to the Maxwell's equations, the force-free condition is
imposed \footnote{In the more conventional form, $\rho_e \mathbf{E}
+ \mathbf{J}\times\mathbf{B} = 0$, where $\rho_e$ and $\mathbf{J}$
are charge and current densities.}
\begin{equation}\label{e:}
 F_{\mu\nu}J^\nu=0.
\end{equation}
In the stationary and axisymmetric case, the equations give rise to
\begin{equation}\label{e:}
 \frac{dA_t}{dA_\phi}=-\om(A_\phi(\rho,\th)).
\end{equation}
That is, the angular velocity $\om$ of the EM fields is a function
of $A_\phi$. Another quantity that is a function of $A_\phi$ is
$B_T$. In the original paper \cite{Blandford:1977ds}, it is defined
by $\hat{B}_T=\sqrt{-\hat{g}}\hat{g}^{rr}\hat{g}^{\th\th}F_{r\th}
=(\triangle/\Si)\sin\th F_{r\th}$. Since the redefinition of $r$
leads to the relation $F_{r\th}=2F_{\rho\th}/(\la\ep r_+\sinh\rho)$,
we have $\hat{B}_T=f(\th)\sinh\rho F_{\rho\th}/(4r_+)$. Here,
following the same definition, $B_T$ on the near-NHEK is
\begin{equation}\label{e:}
 B_T=\sqrt{-g}g^{\rho\rho}g^{\th\th}F_{\rho\th}
=f(\th)\sinh\rho F_{\rho\th}.
\end{equation}
So we get $\hat{B}_T=(\la\ep/4r_+)B_T=(dt/d\hat{t})B_T$.

Thus, the EM fields can be expressed as
\begin{equation}\label{e:}
 E_\rho=-\frac{\bar{\om}}{r_+^2\Ga^2\sinh\rho}\pa_\rho A_\phi,
\textrm{ }\textrm{ }\textrm{ } B_\rho=\frac{1}{r_+^2\sin\th}\pa_\th
A_\phi,
\end{equation}
\begin{equation}\label{e:EMI1}
 E_\th=-\frac{\bar{\om}}{r_+^2\Ga^2\sinh\rho}\pa_\th A_\phi,
\textrm{ }\textrm{ }\textrm{ } B_\th=\frac{-1}{r_+^2\sin\th}\pa_\rho
A_\phi,
\end{equation}
\begin{equation}\label{e:EMI2}
 E_\phi=0,
\textrm{ }\textrm{ }\textrm{ }
B_\phi=\frac{1}{r_+^2\Ga^2}F_{\rho\th}=\frac{B_T}
{r_+^2\sin\th\sinh\rho},
\end{equation}
where
\begin{equation}\label{e:EMI3}
 \bar{\om}\equiv\om+\cosh\rho. \nonumber
\end{equation}
The field components automatically satisfy the degenerate condition
$\textbf{E}\cdot\textbf{B}=0$.  In contrast to the corresponding
components given in \cite{Znajek:1977unknown}, the changes on the
poloidal EM components mainly come from the transformations
\begin{equation}\label{e:omtrans}
 1-a\sin^2\th\hat{\om}\rightarrow\Ga^2,
\textrm{ }\textrm{ }\textrm{ } (r^2+a^2)\hat{\om}-a\rightarrow
\frac{1}{2}r_+\la\ep\bar{\om},
\end{equation}
under $(\hat{t},\hat{\phi})\rightarrow(t,\phi)$ (in the limit
$\la\rightarrow0$). The transformations are direct results of those
of the relevant one-forms: $d\hat{t}-a\sin^2\th
d\hat{\phi}\rightarrow 4r_+\Ga^2dt/\la\ep$ and
$(r^2+a^2)d\hat{\phi}-ad\hat{t}\rightarrow 2r_+^2(d\phi+\cosh\rho
dt)$.

Assembling all the equations and conditions, we finally arrive at a
stream differential equation for the toroidal vector potential
$A_\phi$. For a general stationary and axisymmetric metric, the
stream equation has been given in \cite{Blandford:1977ds}. Inserting
the near-NHEK metric (\ref{e:Kerr3}), we can express the equation as
\begin{equation}\label{e:EqI}
 \frac{1}{F}\pa_i\left(\frac{
\bar{\om}^2F^2-1}{F}\pa^iA_\phi\right)-\bar{\om}
\om'(\pa_iA_\phi)^2=\frac{B_TB_T'}{f^2},
\end{equation}
where $i=(\rho,\th)$, the primes stand for the derivative with
respect to $A_\phi$ and
\begin{equation}\label{e:F}
 F(\rho,\th)=\frac{f}{\sinh\rho}.
\end{equation}
Denoting $x\equiv\cosh\rho$, we can re-express the equation as
\begin{eqnarray}\label{e:EqII}
 (x^2-1)
\left\{f^2[\bar{\om}\pa_x(\bar{\om}\pa_xA_\phi)+\bar{\om}\pa_x
A_\phi]-\pa_x[(x^2-1)\pa_xA_\phi]-f\pa_\th\left(f^{-1}\pa_\th
A_\phi\right)\right\} \nonumber
\\ +f\bar{\om}\pa_\th(f\bar{\om}\pa_\th
A_\phi)=B_TB_T'.
\end{eqnarray}

\subsection{Boundary condition on the horizon}

The field components (\ref{e:EMI1}) and (\ref{e:EMI2}) in the Carter
frame become finite on the horizon by null rotating into the ingoing
frame. This leads to some constraints on these fields on horizon.
From the constraints, the boundary condition of $\hat{B}_T$ on the
horizon is derived by Znajek in \cite{Znajek:1977unknown}:
\begin{equation}\label{e:oriBhor}
 \hat{B}_T(r=r_+,\th)=\frac{\sin\th[\hat{\om}(r_+^2+a^2)-a]}
{\Si(r_+,\th)}\pa_\th \hat{A}_{\hat{\phi}} (r_+,\th).
\end{equation}
As stated above, the EM field components between two coordinate
systems are related with the transformations (\ref{e:omtrans}).
Hence, from the same constraints on the EM fields, we can get the
boundary condition of $B_T$ on near-NHEK
\begin{equation}\label{e:Bhor}
 B_T(\rho=0,\th)=f(\th)(\om+1)\pa_\th A_\phi.
\end{equation}

Here, we show that this condition can actually be derived directly
from the stream differential equation (\ref{e:EqII}). It is easy to
see that, approaching the horizon $\rho\rightarrow0$ or
$x\rightarrow1$, both sides in the second line of the equation are
identical, if the sum of the terms within the braces in the first
line are not singular. This identification exactly gives rise to the
boundary condition of $B_T$ given in Eq.\ (\ref{e:Bhor}). Of course,
from the equation, the toroidal field also admits the condition
$B_T(\rho=0)=-f(\th)(\om+1)\pa_\th A_\phi$.

\subsection{Separation of the stream equation}

In terms of the boundary condition (\ref{e:Bhor}), it is natural to
guess that the solution of $B_T$ away from the horizon could be of
the form:
\begin{equation}\label{e:BT}
 B_T=f(\th)\bar{\om}\pa_\th A_\phi.
\end{equation}
For this solution of $B_T$, the EM field components have the
relation
\begin{equation}\label{e:EthBphi}
 E_\th=-B_\phi.
\end{equation}
Thus, the two sides of the second line of Eq.\ (\ref{e:EqII}) are
equal and the rest part forms the following equation
\begin{equation}\label{e:remeq}
 f^2[\bar{\om}\pa_x(\bar{\om}\pa_xA_\phi)+\bar{\om}\pa_x
A_\phi]-\pa_x[(x^2-1)\pa_xA_\phi]-f\pa_\th\left(f^{-1}\pa_\th
A_\phi\right)=0.
\end{equation}
We only need to consider consistent solutions to this equation,
accompanied with restriction conditions given above.

\subsection{Near-axis solutions}

Here, we only explore the solutions in the range $\th\in[0,\pi/2]$.
The solutions in the range $\th\in[\pi/2,\pi]$ are the same.

From the equation, we can find that it becomes an equation only
relying on the variable $\th$ if $\om\sim x$ and $A_\phi\sim x$. So
we make the general ansatz as follows
\begin{eqnarray}\label{e:ansrel}
 \om=U(\th)+V(\th)x, && A_\phi=b\om^k+c,
\end{eqnarray}
where $b$, $c$ and $k$ are constants.

Let us denote
\begin{equation}\label{e:}
 L_\th^2\equiv-f^{-1}\pa_\th\left(f^{-1}\pa_\th\right).
\end{equation}
Inserting the ansatz (\ref{e:ansrel}) into Eq.\ (\ref{e:remeq}) and
comparing the coefficients of $x^0$, $x$ and $x^2$ respectively, we
can obtain the following three equations
\begin{equation}\label{e:streq1}
 kV^2+2V+(k-1)f^{-2}\frac{V^2}{U^2}+\frac{L_\th^2U}{U}
-(k-1)f^{-2}\frac{(\pa_\th U)^2}{U^2}=0,
\end{equation}
\begin{equation}\label{e:streq2}
 kV(V+1)+\frac{1}{2}(3V+2)-f^{-2}+\frac{1}{2}\left(\frac{L_\th^2U}
{U}+\frac{L_\th^2V}{V}\right)-(k-1)f^{-2}\frac{\pa_\th U\pa_\th
V}{UV}=0,
\end{equation}
\begin{equation}\label{e:streq3}
 k(V+1)^2+V+1-(k+1)f^{-2}+\frac{L_\th^2V}{V}
-(k-1)f^{-2}\frac{(\pa_\th V)^2}{V^2}=0.
\end{equation}
When $k=1$ or $U=\pm V$, the three equations degenerate to two:
(Eq.\ (\ref{e:streq1})+Eq.\ (\ref{e:streq3}))/2=Eq.\
(\ref{e:streq2}). For the special case $U=0$, the first two
equations become trivial and only the last equation about $V$ is
left.

Generically, the equations are non-linear and are hard to derive
full analytical solutions. But it is nice to notice that they can be
simultaneously solved near the axis $\th\sim0$ by the solutions
\begin{equation}\label{e:UV}
 U\simeq\frac{p}{\th^2},
\textrm{ }\textrm{ }\textrm{ }
V\simeq\frac{q}{\th^2},
\end{equation}
where $p$ is arbitrary constant and
\begin{equation}\label{e:q}
 q^2=\frac{4(k+1)}{k} \textrm{ }\textrm{ }\textrm{ }
(k>0 \textrm{ }\textrm{ }\textrm{ or }\textrm{ }
\textrm{ } k<-1).
\end{equation}
To make $A_\phi$ non-singular on the axis, we must choose $k<-1$.

Next, we need to check if $B_T$ is a function of $\om$ or $A_\phi$.
Inserting the above solutions into Eq.\ (\ref{e:BT}), we get
\begin{equation}\label{e:}
 B_T\simeq-2bk\om^{k+1}.
\end{equation}
So for $k<-1$, $B_T$ is also non-singular on the axis.

From the solution of $A_\phi$ and the relation (\ref{e:EthBphi}),
the scalars on the horizon satisfy
\begin{equation}\label{e:}
 \phi_0(\rho=0)=B_\phi, \textrm{ }\textrm{ }\textrm{ }
\phi_2(\rho=0)=0.
\end{equation}
This describes ingoing flux that is not scattered by the black hole
\cite{Brennan:2013jla,Gralla:2014yja}. But, away from the horizon
with $\rho\neq0$, the imaginary part of $\phi_2$ becomes
non-vanishing. If $B_T$ takes the form $B_T=-f(\th)\bar{\om}\pa_\th
A_\phi$ (which is allowed by the boundary condition on the horizon
as stated in the previous subsection), then the relation in Eq.\
(\ref{e:EthBphi}) becomes $E_\th=B_\phi$ and so $\phi_0(\rho=0)=0$,
$\phi_2(\rho=0)=B_\phi$, which describes outgoing flux. The two flux
modes can be reversed by the simultaneous interchanges of the
coordinates $t\leftrightarrow-t$ and $\phi\leftrightarrow-\phi$.

\subsection{Comparison with the MD solutions}

The solutions (\ref{e:ansrel}) with (\ref{e:UV}) are consistent with
those obtained under far-field approximation in
\cite{Menon:2005va,Menon:2005mg,Menon:2011zu}. In there, two sets of
solutions are derived in the asymptotic region, where the fields and
quantities become radial distance independent. In the original BL
coordinates, one set of the solutions read \cite{Menon:2011zu}
\begin{equation}\label{e:MD1}
 \Om_-=\frac{1}{a\sin^2\th}, \textrm{ }\textrm{ }\textrm{ }
H_\varphi=\frac{2\La\cos\th}{a^2}\frac{1}{\sin^4\th},
\end{equation}
\begin{equation}\label{e:MD2}
 E_\th=-\frac{2\La\cos\th}{a^2}\frac{1}{\sin^5\th},
\textrm{ }\textrm{ }\textrm{ }
B_r=\frac{2\sqrt{\triangle}\La\cos\th}{a\Si\chi}
\frac{1}{\sin^4\th},
\end{equation}
where $\La(\th)$ is an arbitrary function about $\th$ and
$\chi^2=(r^2+a^2)^2-\triangle a^2\sin^2\th$.

For our near-horizon and near-axis solutions, the corresponding
quantities are
\begin{equation}\label{e:}
 \om\simeq\frac{p+qx}{\th^2}, \textrm{ }\textrm{ }\textrm{ }
B_T\simeq-2bk\om^{k-1}\frac{(p+qx)^2}{\th^4},
\end{equation}
\begin{equation}\label{e:}
 E_\th\simeq\frac{2bk\om^{k-1}}{r_+^2\sinh\rho}
\frac{(p+qx)^2}{\th^5}, \textrm{ }\textrm{ }\textrm{ }
B_\rho\simeq-\frac{2bk\om^{k-1}}{r_+^2}\frac{p+qx}{\th^4}.
\end{equation}
Thus, the $\th$-dependent parts of the solutions are consistent with
the MD solutions (\ref{e:MD1}) and (\ref{e:MD2}) around $\th\sim0$,
if $\om^{k-1}$ has the same $\th$ dependence as $-\La(\th)$. As done
in \cite{Menon:2005mg,Menon:2011zu}, the singular behaviour on the
poles can be avoided by some transformations like
$\La\rightarrow\La\sin^5\th$. In our situation, this corresponds to
$k=-3/2$.

%%%%%%%%%%%%%%%%%%%%%%%%%%%%%%%%%%%%%%%%%%%%%%%%%%%%%%%%%%%%%%%%%%%%%%%%%%%%
\section{Energy and angular-momentum extraction}
\label{sec:extraction}
%%%%%%%%%%%%%%%%%%%%%%%%%%%%%%%%%%%%%%%%%%%%%%%%%%%%%%%%%%%%%%%%%%%%%%%%%%%%

Now, let us check whether the derived solutions can lead to jet-like
outflow formation. As above, we denote the quantities on the
original BL coordinates (\ref{e:Kerr}) by a hat. The vector
potentials in the two coordinate systems are related via
\begin{equation}\label{e:}
 \hat{A}_{\hat{t}}=\frac{1}{4r_+}\left(\la\ep A_t
-2A_\phi\right), \textrm{ }\textrm{ }\textrm{ }
\hat{A}_{\hat{\phi}}=A_\phi.
\end{equation}
The relation between the angular velocities is
\begin{equation}\label{e:}
 \hat{\om}=\left(1+\frac{1}{2}\la\ep\om\right)\Om_H,
\end{equation}
where $\Om_H\equiv 1/(2r_+)$. Towards the horizon $\la\rightarrow0$,
$\hat{\om}\rightarrow \Om_H$ as expected. So it seems that we have
done an expansion of $\hat{\om}$ near the horizon to the order
$\mathcal{O}(\la)$ in the above discussion.

As given in Eq.\ (\ref{e:omtrans}), $(r^2+a^2)\hat{\om}-a\rightarrow
2r_+^2\bar{\om}(dt/d\hat{t})$. In terms of the boundary condition
(\ref{e:oriBhor}) and the solution (\ref{e:BT}), we can express the
solution of $\hat{B}_{T}$ in the near-horizon region as:
\begin{equation}\label{e:}
 \hat{B}_{T}(r,\th)=\frac{\sin\th[\hat{\om}(r^2+a^2)-a]}
{\Si(r,\th)}\pa_\th \hat{A}_{\hat{\phi}} (r,\th) \simeq
B_T(\rho,\th)\frac{dt}{d\hat{t}},
\end{equation}
where $r$ is closed to $r_+$.

Let us say that the near-axis solutions obtained in the previous
section are valid only within $0\leq\th\leq\th_c$, with $\th_c$
representing the critical value where the approximation is
appropriate. Then, the rate of energy extraction within this range
is
\begin{equation}\label{e:}
 \frac{d\hat{E}}{d\hat{t}}=-\int_0^{2\pi} d\hat{\phi}
\int_{0}^{\th_c}d\th\hat{\om}\hat{B}_T
\pa_\th\hat{A}_{\hat{\phi}}\simeq-\la\ep
b^2k^2\Om_H^2\left(\frac{\left[\om^{2k+1}
\right]_{0}^{\th_c}}{2k+1}+\frac{\la\ep\left[
\om^{2(k+1)}\right]_{0}^{\th_c}}{4(k+1)}\right),
\end{equation}
which is positive for $k<-1$ (as well as $p,q>0$). Similarly, the
rate of angular momentum extraction is
\begin{equation}\label{e:}
 \frac{d\hat{L}}{d\hat{t}}\simeq-\frac{\la\ep b^2k^2\Om_H}{2k+1}
\left[\om^{2k+1} \right]_{0}^{\th_c}.
\end{equation}
It is also positive for $k<-1$. So there exists outward flux of
energy and angular-momentum from the hole.

%%%%%%%%%%%%%%%%%%%%%%%%%%%%%%%%%%%%%%%%%%%%%%%%%%%%%%%%%%%%%%%%%%%%%%%%%%%%
\section{Conclusions}
\label{sec:conclusion}
%%%%%%%%%%%%%%%%%%%%%%%%%%%%%%%%%%%%%%%%%%%%%%%%%%%%%%%%%%%%%%%%%%%%%%%%%%%%

It is possible that the (near-)NHEK makes a good tool to
analytically study the force-free magnetosphere near rapidly
rotating astronomical black holes. By focusing on the near-horizon
region, we showed that the boundary condition on the horizon can be
derived directly from the stream equation on near-NHEK. We got
consistent solutions near the rotation axis which imply the
formation of outflow. The angular dependence of the solutions
resembles that of the asymptotic solutions obtained in
\cite{Menon:2005va,Menon:2005mg,Menon:2011zu}.

%\section*{Acknowledgements\markboth{Acknowledgements}{Acknowledgements}}

%\noindent

%\newpage
\bibliographystyle{JHEP}
\bibliography{b}

\end{document}